\documentclass[aps,prl,twocolumn,10pt,superscriptaddress]{revtex4-1}

\setcitestyle{super}

\usepackage[pdftex]{graphicx}
\usepackage{epstopdf}
\usepackage{amsmath}
\usepackage{amssymb}
\usepackage{bm}

\renewcommand{\figurename}{\textbf{Figure}}

\begin{document}
\title{Parametric resonance of magnetization excited by electric field}

\author{Yu-Jin~Chen}
\thanks{These authors contributed equally to this work.}
\author{Han~Kyu~Lee}
\thanks{These authors contributed equally to this work.}
\affiliation{Department of Physics and Astronomy, University of California, Irvine, California 92697, USA}
\author{Roman~Verba}
\thanks{These authors contributed equally to this work.}
\affiliation{Institute of Magnetism, Kyiv 03142, Ukraine}
\author{Jordan~A.~Katine}
\affiliation{HGST, San Jose, California 95135, USA}
\author{Igor~Barsukov}
\affiliation{Department of Physics and Astronomy, University of California, Irvine, California 92697, USA}
\author{Vasil~Tiberkevich}
\affiliation{Department of Physics, Oakland University, Rochester, Michigan 48309, USA}
\author{John~Q.~Xiao}
\affiliation{Department of Physics and Astronomy, University of Delaware, Newark, Delaware 19716, USA}
\author{Andrei~N.~Slavin}
\affiliation{Department of Physics, Oakland University, Rochester, Michigan 48309, USA}
\author{Ilya~N.~Krivorotov}
\affiliation{Department of Physics and Astronomy, University of California, Irvine, California 92697, USA}

\begin{abstract}
Manipulation of magnetization by electric field is a central goal of spintronics because it enables energy-efficient operation of spin-based devices. Spin wave devices are promising candidates for low-power information processing but a method for energy-efficient excitation of short-wavelength spin waves has been lacking.  Here we show that spin waves in nanoscale magnetic tunnel junctions can be generated via parametric resonance induced by electric field. Parametric excitation of magnetization is a versatile method of short-wavelength spin wave generation, and thus our results pave the way towards energy-efficient nanomagnonic devices.
\end{abstract}

\maketitle

\section*{Introduction}

Magneto-electric coupling in magnetic materials and heterostructures enables control of magnetization by electric field, which is the key requirement for realization of energy-efficient spintronic devices \cite{Matsukura2015}.  Recent progress in this field includes demonstration of electric field induced magnetization reversal \cite{Chiba2003,Baek2010,Shiota2012,Cuellar2014,Heron2011,He2010,Bauer2015} and ferromagnetic resonance \cite{Nozaki2012,Zhu2012}. However, ferromagnetic resonance driven by electric field cannot be used for generation of spin waves with wavelengths smaller than the excitation region, which limits its applicability in nanomagnonic devices based on short-wavelength spin waves \cite{Yu2016}. Here we report parametric excitation of spin waves in a ferromagnet by alternating electric field. Unlike ferromagnetic resonance, parametric resonance can be employed for generation and amplification of short-wavelength spin waves and thus our work is an important step towards the development of energy-efficient nanomagnonics.

A prominent manifestation of the magneto-electric coupling in magnetic films and heterostructures is modification of magnetic anisotropy by electric field \cite{Weisheit2007,Chiba2008,Maruyama2009}. This recently discovered effect takes place at the interface between a ferromagnetic metal (e.g.~Fe) and a nonmagnetic insulator (e.g.~MgO) \cite{Maruyama2009} and originates from different rates of filling of $\it{d}$-like electron bands in response to electric field applied perpendicular to the interface \cite{Niranjan2010}. Since electrons in different bands contribute unequally to the uniaxial perpendicular magnetic anisotropy (PMA) at the interface, electric field can be used to modulate PMA. This voltage-controlled magnetic anisotropy (VCMA) is promising for energy-efficient manipulation of magnetization \cite{Shiota2012,Wang2012,Schellekens2012} because, unlike spin torque (ST), VCMA does not rely on high electric current density resulting in large Ohmic losses. In this work, we employ VCMA modulation at microwave frequencies in order to excite parametric resonance of magnetization in a nanomagnet \cite{Verba2014,Kurebayashi2011,Manuilov2015}.

Parametric excitation of magnetization by external magnetic field has been thoroughly studied in bulk and thin-film ferromagnets \cite{Gurevich1996}. In these experiments, a parameter of the magnetic system (external field) is modulated with a frequency at twice a spin wave frequency $f_\mathrm{SW}$ of the system.  Parametric excitation is a nonlinear process, in which the parametric drive acts as negative effective magnetic damping competing with positive intrinsic damping \cite{Gurevich1996}. At a threshold amplitude of the parametric drive, the negative damping exceeds the intrinsic damping and magnetization oscillations at half the drive frequency are excited. 

Parametric excitation of magnetization has several important advantages over direct excitation by external magnetic field with a frequency at $f_\mathrm{SW}$. First, parametric excitation efficiently couples not only to the uniform precession of magnetization but also to spin wave eigenmodes \cite{Gurevich1996}. This allows excitation of short wavelength spin waves by simply choosing the parametric drive frequency to be twice the desired spin wave frequency. Second, parametric pumping can be used for frequency-selective amplification of spin waves \cite{Melkov2001} and phase error corrections \cite{Chumak2012}. All these properties of parametric pumping form a highly desirable set of tools for the nascent field of nanomagnonics \cite{Kruglyak2010,Chumak2015}. However, parametric excitation of spin waves by microwave magnetic field in metallic ferromagnets is not energy-efficient because of the relatively high threshold fields (tens of Oe) \cite{Urazhdin2010}. Here we show that replacing magnetic field pumping by electric field (VCMA) pumping solves this problem and allows parametric excitation of magnetic oscillations in metallic ferromagnets by a low-power microwave drive.

\begin{figure*}[ht]
\includegraphics[width=2.0\columnwidth]{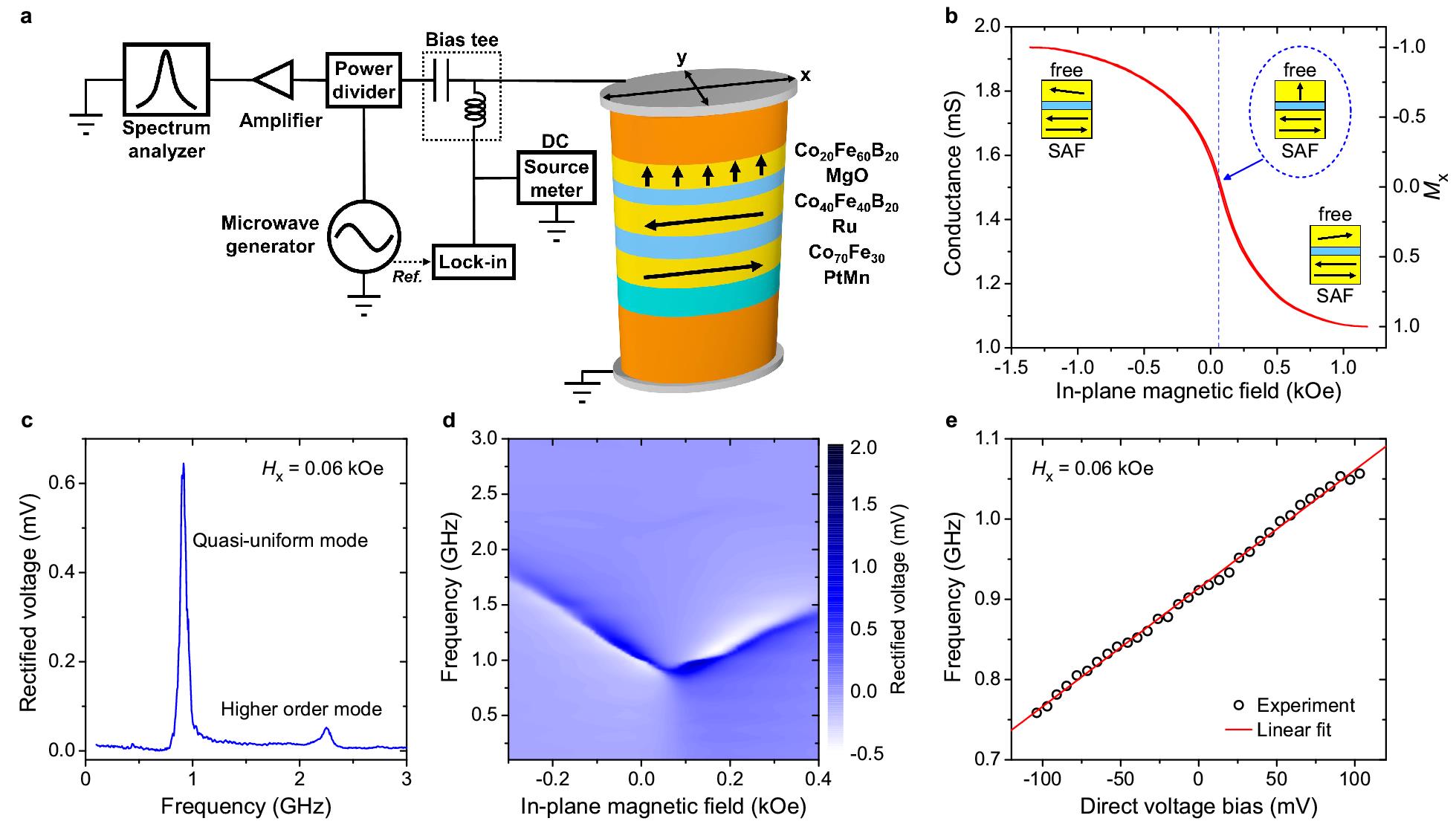}
\caption{\textbf{Measurement setup and MTJ characterization.} \textbf{a}, Schematic of experimental setup for DC and microwave characterization of MTJ. \textbf{b}, MTJ conductance as a function of in-plane magnetic field $H_\mathrm{x}$ applied parallel to the MTJ long axis. \textbf{c}, ST-FMR spectrum of the MTJ at $H_\mathrm{x}=0.06$\,kOe. \textbf{d}, Dependence of ST-FMR spectra on $H_\mathrm{x}$. \textbf{e}, Quasi-uniform mode frequency versus direct voltage bias $V_\mathrm{dc}$ measured at $H_\mathrm{x} = 0.06$\,kOe.}\label{f:fmr}
\end{figure*}

We demonstrate parametric excitation of magnetization in 70\,nm$\times$150\,nm elliptical nanoscale magnetic tunnel junctions (MTJs) schematically shown in Fig.\,\ref{f:fmr}a. The junctions are patterned from (bottom lead)/ Ta(5)/ PtMn(15)/ SAF/ MgO(0.83)/ Co$_{20}$Fe$_{60}$B$_{20}$(1.58)/ Ta(5)/ (cap) multilayers (thicknesses in nm) deposited by magnetron sputtering. Here SAF = Co$_{70}$Fe$_{30}$(2.3)/ Ru(0.85)/ Co$_{40}$Fe$_{40}$B$_{20}$(2.4) is the pinned synthetic antiferromagnet, which has magnetic moments lying in the plane of the sample. The equilibrium direction of the Co$_{20}$Fe$_{60}$B$_{20}$ free layer magnetization is normal to the sample plane due to interfacial PMA \cite{Zhu2012}.  Prior to patterning, the multilayers are annealed for 2 hours at 300\,$^\circ$C in a 10\,kOe in-plane magnetic field that sets the pinned layer exchange bias direction parallel to the MTJ long axis. 

All measurements reported in this Article are made in the setup schematically shown in Fig.\,\ref{f:fmr}a that allows application of DC and microwave voltages to the MTJ and measurement of DC and microwave signals generated by the MTJ. Fig.\,\ref{f:fmr}b shows conductance $G$ of the MTJ measured as a function of in-plane magnetic field $H_\mathrm{x}$ applied parallel to the MTJ long axis. The shape of the $G(H_\mathrm{x})$ curve is congruent to the shape of the $M_\mathrm{x}(H_\mathrm{x})$ hysteresis loop \cite{Zhu2012}, where $M_\mathrm{x}$ is normalized projection of the free layer magnetization onto the applied field direction. The hysteresis loop and micromagnetic simulations confirm the out-of-plane easy axis of the free layer (see Supplementary Section 1). The center of the loop is shifted from zero field due to a residual 0.06\,kOe stray field from the SAF. 

We employ spin torque ferromagnetic resonance (ST-FMR) to characterize the spectral properties of spin wave eigenmodes of the MTJ. In this technique, a small amplitude microwave drive current $G V_\mathrm{ac} \sin(2\pi f_\mathrm{d} t)$ applied to the MTJ excites oscillations of magnetization at the drive frequency $f_\mathrm{d}$. The resulting resistance oscillations $R_\mathrm{ac} \sin(2\pi f_\mathrm{d} t+ \phi)$ of the MTJ at the drive frequency lead to partial rectification of the microwave drive voltage $V_\mathrm{ac}$ and generate a direct voltage $V_\mathrm{r}$. Peaks in ST-FMR spectra $V_\mathrm{r}(f_\mathrm{d})$ arise from resonant excitation of spin wave eigenmodes of the MTJ \cite{Tulapurkar2005,Sankey2006}. 

Figure\,\ref{f:fmr}c shows a ST-FMR spectrum of the MTJ measured at $H_\mathrm{x} = 0.06$\,kOe. Two spin wave eigenmodes are present in this spectrum with the lowest-frequency ($f_\mathrm{SW} = 0.91$\,GHz) mode being the quasi-uniform mode of the free layer \cite{Goncalves2013}. From the spectral linewidth of the quasi-uniform mode we can estimate the Gilbert damping parameter $\alpha \approx 0.033$ (see Supplementary Section 2), which is typical for a CoFeB layer of this thickness \cite{Zhu2012}. Dependence of ST-FMR spectra on $H_\mathrm{x}$ is summarized in Fig.\,\ref{f:fmr}d. The frequency of the quasi-uniform mode increases with increasing absolute value of the net in-plane field due to the second-order uniaxial PMA \cite{Zhu2012}. 

Figure\,\ref{f:fmr}e shows dependence of the quasi-uniform mode frequency on direct voltage bias $V_\mathrm{dc}$ applied to the MTJ.  From the slope of the line in Fig.\,\ref{f:fmr}e we can estimate VCMA efficiency $\frac{d H_\mathrm{u}}{d V_\mathrm{dc}}=526$\,Oe/V (see Supplementary Section 3), where $H_\mathrm{u}$ is the PMA field and the value of VCMA efficiency is typical for this material system \cite{Zhu2012}.

\begin{figure*}[ht]
\includegraphics[width=2.0\columnwidth]{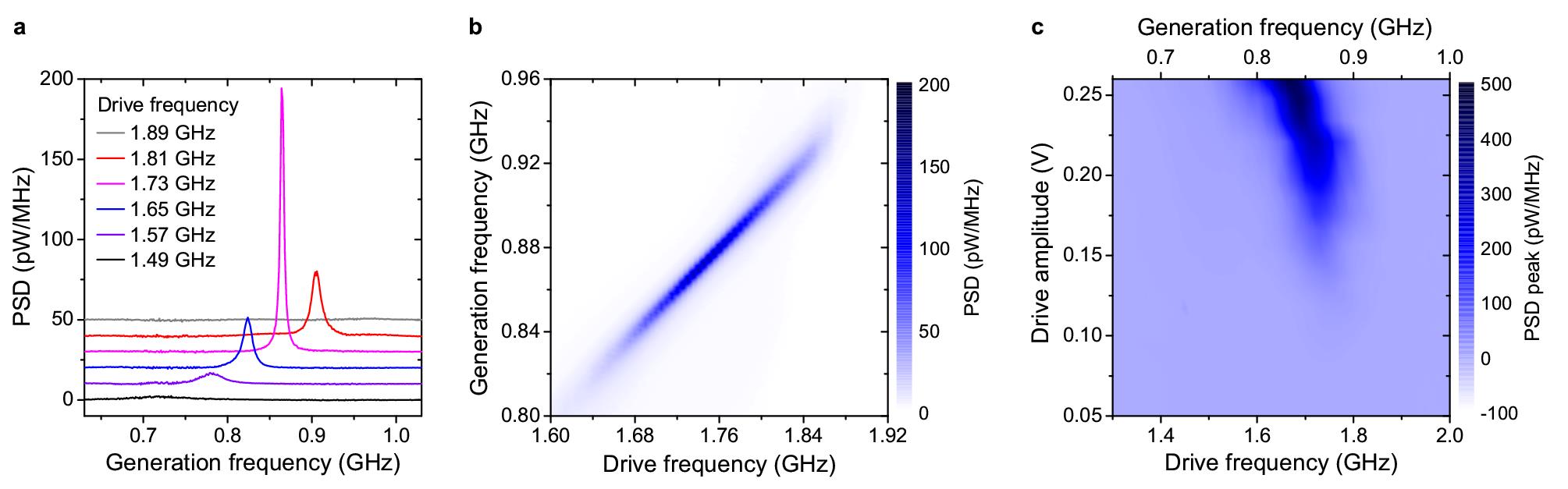}
\caption{\label{fig:schematic}\textbf{Parametric resonance.} \textbf{a}, Power spectral density (PSD) of the microwave signal emitted by the MTJ under VCMA parametric drive of $V_\mathrm{ac} = 0.185$\,V. Curves are vertically offset for clarity and are listed in order of drive frequency. \textbf{b}, Dependence of the parametrically generated emission spectra on the drive frequency for $V_\mathrm{ac} = 0.185$\,V. \textbf{c}, PSD peak plotted versus drive frequency and drive amplitude reveals typical Arnold tongue shape characteristic of parametric excitation.} \label{f:pr}
\end{figure*}

\section*{Results}

We use the parallel pumping geometry to parametrically excite the free layer quasi-uniform mode \cite{Verba2014}, in which magnetization of the free layer is parallel to the oscillating PMA field $H_\mathrm{u}$. We apply a constant 0.06\,kOe in-plane magnetic field along the long axis of the ellipse to compensate the in-plane SAF stray field acting on the free layer so that its magnetization is aligned perpendicular to the sample plane. We then apply a parametric drive voltage $V_\mathrm{ac}$ to the MTJ and vary the drive frequency $f_\mathrm{d}$ about $2 f_\mathrm{SW}$ (twice the quasi-uniform mode resonance frequency). The resulting modulation of PMA at the drive frequency due to VCMA can parametrically excite magnetization oscillations at half the drive frequency \cite{Verba2014}, which gives rise to the MTJ resistance oscillations $R_\mathrm{ac}\cos(2\pi\frac{f_\mathrm{d}}{2}t+\phi)$. These resistance oscillations can be detected via their mixing with the microwave current $G V_\mathrm{ac}\cos(2\pi f_\mathrm{d} t)$ through the junction, which generates mixing voltage signals $V_{\mathrm{mix}} (t)$ at frequencies $f_\mathrm{d}/2$ and $3 f_\mathrm{d}/2$:
\begin{eqnarray}
 \label{eq:1}
  V_{\mathrm{mix}} (t) = G V_\mathrm{ac}\cos( 2 \pi f_\mathrm{d} t) \cdot R_\mathrm{ac}\cos\bigg(2 \pi \frac{f_\mathrm{d}}{2} t + \phi\bigg) = \nonumber \\
\frac{1}{2} G V_\mathrm{ac}R_\mathrm{ac} \Bigg[\cos\bigg(2 \pi \frac{f_\mathrm{d}}{2} t - \phi\bigg) + \cos\bigg(2 \pi \frac{3f_\mathrm{d}}{2} t + \phi\bigg) \Bigg].
\end{eqnarray}

As illustrated in Fig.\,\ref{f:fmr}a, we amplify $V_\mathrm{mix}(t)$ and measure its spectrum with a microwave spectrum analyzer. In this Article, we present power spectra of $V_\mathrm{mix}(t)$ measured near $f_\mathrm{d}/2$; similar spectra are observed near $3 f_\mathrm{d}/2$. Figure\,\ref{f:pr}a displays power spectral density (PSD) $P(f)$ of $V_\mathrm{mix}(t)$ measured at several fixed values of the drive frequency $f_\mathrm{d}$ near $2 f_\mathrm{SW}$ and drive amplitude $V_\mathrm{ac}= 0.185$\,V. The maximum of each power spectrum is observed exactly at $f_\mathrm{d}/2$, clearly illustrating that magnetization dynamics of the free layer is excited parametrically at half the drive frequency. The linewidths of the measured spectral peaks are in the range of several MHz. This linewidth mostly arises from thermal fluctuations of the free layer magnetization (fluctuations of the phase $\phi$ in equation\,(\ref{eq:1})). Figure\,\ref{f:pr}b illustrates that parametric excitation of the quasi-uniform mode has well-pronounced resonant character: significant amplitude of the parametric oscillations is observed only in a narrow range of the drive frequencies near $2 f_\mathrm{SW}$.

Figure\,\ref{f:pr}c displays dependence of $P(f_\mathrm{d}/2)$ on the drive amplitude $V_\mathrm{ac}$ and drive frequency $f_\mathrm{d}$. This figure illustrates the parametric excitation efficiency and clearly demonstrates that the observed microwave emission from the sample has a threshold character in $V_\mathrm{ac}$. This threshold behavior is expected for parametric resonance that is excited when effective negative damping from the parametric drive exceeds the positive natural damping of the excited mode \cite{Gurevich1996}. Figure\,\ref{f:pr}c also shows that the parametric resonance frequency $f_\mathrm{pr}$ (defined as $f_\mathrm{d}$ that gives maximum $P(f_\mathrm{d}/2)$ at a given value of $V_\mathrm{ac}$) shifts to lower values with increasing drive amplitude due to nonlinear frequency shift, as expected for a uniaxial ferromagnet \cite{Gurevich1996}. The shape of the parametric instability region in Fig.\,\ref{f:pr}c is a typical Arnold tongue of a nonlinear parametric oscillator \cite{Bortolotti2014}.

\begin{figure}[ht]
\includegraphics[width=0.95\columnwidth]{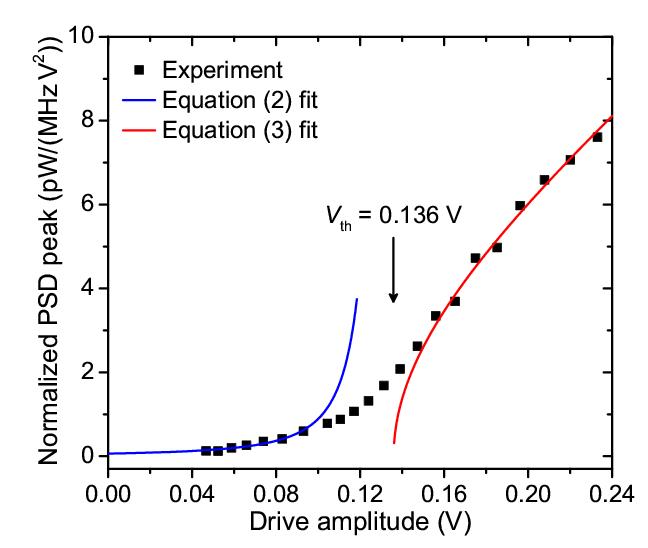}
\caption{\label{fig:prthreshold}\textbf{Parametric resonance threshold.} Normalized peak amplitude of PSD, $P(f_\mathrm{pr}/2)/V_\mathrm{ac}^2$, measured at parametric resonance as a function of the parametric drive amplitude $V_\mathrm{ac}$. Best fits of equation\,(\ref{eq:low}) and equation\,(\ref{eq:high}) to the data (solid lines) give the parametric resonance threshold voltage $V_\mathrm{th}$ = 0.136\,V.} \label{f:th}
\end{figure}

In order to quantitatively determine the threshold drive voltage $V_\mathrm{th}$ needed to excite parametric resonance of the quasi-uniform mode, we analyze reduced power of this mode $p$ as a function of the drive amplitude $V_\mathrm{ac}$. By definition, $p = |c|^2$ where $c$ is dimensionless amplitude of the quasi-uniform mode (see Supplementary Section 4), which is proportional to the amplitude of the MTJ resistance oscillations, so that $p \sim (G R_\mathrm{ac})^2$. It is clear from equation\,(\ref{eq:1}) that PSD of the reduced power $p(f)$ is proportional to  $P(f)/V_\mathrm{ac}^2$ for any $V_\mathrm{ac}$. In Fig.\,\ref{f:th}, we plot its resonant value $P(f_\mathrm{pr}/2)/V_\mathrm{ac}^2$, which is proportional to $p(f_\mathrm{pr}/2)$, as a function of $V_\mathrm{ac}$. 

Analytical expressions for $p(f_\mathrm{pr}/2)$  can be derived in the limit of $V_\mathrm{ac}\ll V_\mathrm{th}$. In this limit, magnetization dynamics are small-amplitude thermal fluctuations amplified by the parametric drive, for which:
\begin{equation}
 \label{eq:low}
	p(f_\mathrm{pr}/2) = \frac{A}{(V_\mathrm{th} - V_\mathrm{ac})^2},
\end{equation}
where $A$ is a constant (see Supplementary Section 5).

In the opposite limit of $V_\mathrm{ac}\gg V_\mathrm{th}$, thermal fluctuations can be neglected and the following analytical expression for the reduced power $p$ can be derived:
\begin{equation}
 \label{eq:high}
	p = B \sqrt{V_\mathrm{ac}^2 - V_\mathrm{th}^2},
\end{equation}
where $B$ is a constant (see Supplementary Section 5). 

For our system, $p$ in equation\,(\ref{eq:high}) can be replaced by $p(f_\mathrm{pr}/2)$ because the measured spectral linewidth of $P(f)$ at $f_\mathrm{d}=f_\mathrm{pr}$ depends weakly on $V_\mathrm{ac}$ for $V_\mathrm{ac} > 0.16$\,V. Therefore, we can fit the data in Fig.\,\ref{f:th} using equation\,(\ref{eq:low}) in the small amplitude limit and equation\,(\ref{eq:high}) in the large amplitude limit. 
The best fit shown by the blue (small amplitude) and red (large amplitude) lines in Fig.\,\ref{f:th} gives $V_\mathrm{th}=0.136$\,V. In this fitting procedure, $A$ and $B$ are free fitting parameters while $V_\mathrm{th}$ is treated as a common fitting parameter for both the small and large amplitude limits (see Supplementary Section 5).

\begin{figure}[ht]
\includegraphics[width=0.95\columnwidth]{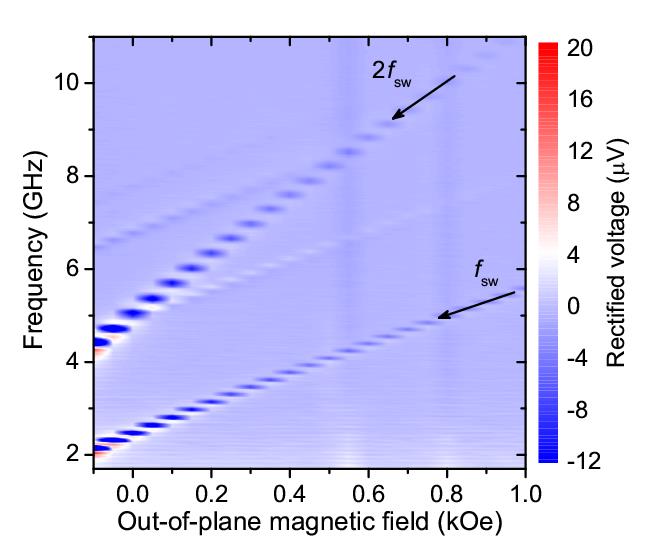}
\caption{\label{fig:prst-fmr}\textbf{Parametric resonance in ST-FMR.} ST-FMR spectra of an MTJ with out-of-plane SAF and free layers measured as a function of out-of-plane magnetic field. Resonance at twice the quasi-uniform mode frequency arises from parametric excitation of the quasi-uniform mode.} \label{f:st-fmr}
\end{figure}

It is instructive to compare the measured $V_\mathrm{th}$ to its theoretically expected value for our MTJ geometry and the measured VCMA efficiency (see Supplementary Section 4). The calculated threshold voltage in the macrospin approximation is $V_\mathrm{th}= 0.086$\,V  while that given by micromagnetic simulations is $V_\mathrm{th}= 0.156$\,V. The measured value is similar to the micromagnetic prediction, which lends support to VCMA origin of the observed parametric resonance. 

In our experiment, spin-polarized tunneling current flows through the MTJ, which results in ST and Oersted field acting on the free layer. However, these types of drive play a negligible role in exciting parametric resonance compared to the VCMA drive. The Oersted field has nearly circular symmetry and therefore it poorly couples to the quasi-uniform mode. The effective fields of both the field-like and damping-like ST lie in the sample plane, which corresponds to perpendicular pumping geometry. Parametric excitation of the quasi-uniform mode at $f_d=2f_\mathrm{SW}$ is inefficient in this geometry \cite{Gurevich1996}.

Our experiment employs MTJ magnetic configuration with in-plane SAF and out-of-plane free layer that is convenient for unambiguous demonstration and quantitative analysis of parametric resonance excited by VCMA. However, we find that VCMA-driven parametric resonance can be observed in other types of MTJ configurations. Figure\,\ref{f:st-fmr} shows out-of-plane magnetic field dependence of ST-FMR spectra measured for a 30\,nm$\times$95\,nm MTJ with out-of-plane equilibrium configuration of both the free and SAF layers. Owing to the smaller amplitude of the rectified voltage in this collinear geometry, we employ ultra-sensitive ST-FMR with magnetic field modulation \cite{Goncalves2013} rather than conventional ST-FMR with amplitude modulation. 

The ST-FMR spectra measured at a large value of the microwave drive voltage $V_\mathrm{ac} = 0.4$\,V reveal several spin wave eigenmodes of the free layer. Another prominent resonance is observed at twice the frequency of the lowest-frequency (quasi-uniform) spin wave eigenmode. In this collinear MTJ geometry, the microwave resistance oscillations of the device have a significant component at twice the excited spin wave mode frequency and mix with the parametric drive at twice the mode frequency to give rise to a rectified voltage peak at $2f_\mathrm{SW}$ measured by ST-FMR. The amplitude of this additional resonance at $2 f_\mathrm{SW}$ relative to the amplitude of the resonance at $f_\mathrm{SW}$ increases with increasing $V_\mathrm{ac}$, which is a signature of a thermally smeared threshold behavior similar to that in Fig.\,\ref{f:th}. The out-of-plane collinear geometry is commonly employed in ST magnetic memory (STT-MRAM), and parametric resonance signals in ST-FMR of STT-MRAM can potentially be used for characterization of the free layer properties such as magnetic damping. 

In summary, our work shows that magneto-electric coupling can be used to excite parametric resonance of magnetization by electric field. We employ voltage-controlled magnetic anisotropy at the CoFeB/MgO interface to excite parametric oscillations of a CoFeB free layer magnetization in nanoscale magnetic tunnel junctions. The threshold voltage for parametric excitation in this system is found to be well below 1\,Volt, which is attractive for applications in energy-efficient spintronic and magnonic nanodevices such as spin wave logic \cite{Khitun2010}. This work opens a new energy-efficient route for excitation of magnetization dynamics in thin films of metallic ferromagnets and nanodevices based on magnetic multilayers.

\section*{Acknowledgments} 
We acknowledge the Center for NanoFerroic Devices (CNFD) and the Nanoelectronics Research Initiative (NRI) for funding of this work. This work was also supported by NSF Grants ECCS-1309416, ECCS-1305586, DMR-1610146, by DTRA Grant HDTRA1-16-1-0025 and by the FAME Center, one of six centers of STARnet, an SRC program sponsored by MARCO and DARPA. The authors declare no competing financial interests.

\clearpage
\widetext

\begin{center}
\textbf{\large Supplementary Information: Parametric resonance of magnetization excited by electric field}
\bigskip
\end{center}
\onecolumngrid

\setcounter{equation}{0}
\setcounter{figure}{0}
\setcounter{table}{0}
\setcounter{page}{1}

\renewcommand{\bibnumfmt}[1]{[S#1]}
\renewcommand{\citenumfont}[1]{S#1}

\renewcommand{\thesection}{S\arabic{section}}
\renewcommand{\thesubsection}{S\arabic{section}\Alph{subsection}}
\renewcommand{\theequation}{S\arabic{equation}}

\renewcommand{\figurename}{\textbf{Supplementary Figure}}

\section{Micromagnetic simulations}
\begin{figure}[h]
\includegraphics[width=0.75\textwidth]{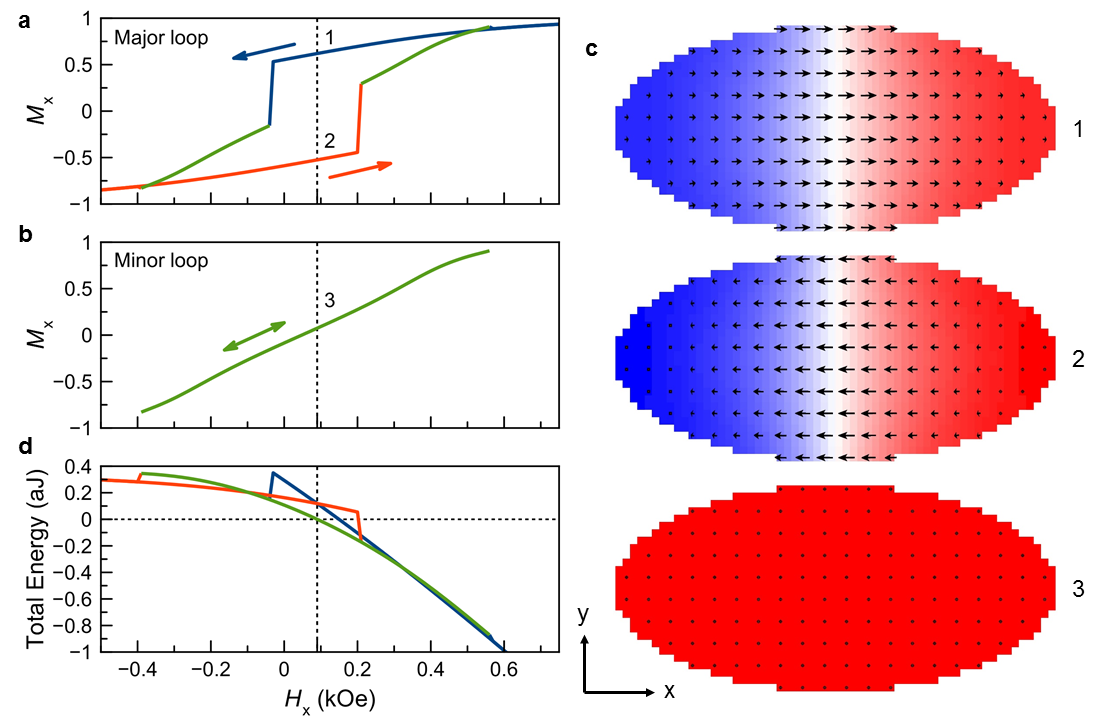}
\caption{\textbf{Micromagnetic simulations.} \textbf{a},\textbf{b}, Normalized magnetization of the free layer $M_\mathrm{x}$ as a function of in-plane magnetic field $H_\mathrm{x}$ applied parallel to the MTJ long axis, given by zero-temperature micromagnetic simulations (\textbf{a} - major loop of the free layer, \textbf{b} - minor loop of the free layer). Arrows indicate field sweep direction. \textbf{c}, Three stable micromagnetic states of the free layer in the middle of the hysteresis loop: two N\'{e}el domain wall states with opposite directions of magnetization in the middle of the domain wall (1 and 2) and the quasi-uniform state (3). Arrows show the in-plane magnetization component while the colors represent the out-of-plane magnetization component with red (blue) being positive (negative). \textbf{d}, Total micromagnetic energy of the MTJ as a function of $H_\mathrm{x}$. Colors correspond to the free layer micromagnetic states shown in \textbf{a}, \textbf{b}, and \textbf{c}. }
\label{f:sim_full_loop}
\end{figure}

To determine the magnetic ground state of the MTJ nanopillar studied in this work, we perform  micromagnetic simulations of the entire nanopillar stack (including the free and the SAF layers) using OOMMF software \cite{Donahue1999}. The magnetic material parameters of the free and the SAF layers employed in the simulations were determined in previous studies of these MTJ devices \cite{Zhu2012, Goncalves2013}. We simulate the hysteresis loops of the device at zero temperature as a function of in-plane magnetic field $H_\mathrm{x}$. Supplemental~Figure\,\ref{f:sim_full_loop}a shows the major loop of the normalized free layer magnetization $M_\mathrm{x}$ as a function of $H_\mathrm{x}$ (the SAF layer magnetization does not switch for the range of $H_\mathrm{x}$ employed). This loop is shifted from zero field by a value similar to that observed in the experiment (see Fig.\,1b of the main text) due to stray magnetic field from the SAF layer acting on the free layer. 

The simulations reveal the presence of three stable states of the free layer magnetization at low fields as illustrated in Supplemental Fig.\,\ref{f:sim_full_loop}c: two N\'{e}el domain wall states with opposite directions of magnetization in the middle of the domain wall (1 and 2) and the quasi-uniform state (3). A minor hysteresis loop shown in Supplemental Fig.\,\ref{f:sim_full_loop}b demonstrates that the quasi-uniform state of the free layer is stable in a wide range of magnetic fields near zero, suggesting that this state is the ground state near zero field. This is directly confirmed by plotting the total micromagnetic energy of the MTJ as a function of $H_\mathrm{x}$ for all three micromagnetic states of the system (Supplemental Fig.\,\ref{f:sim_full_loop}d). This plot clearly shows that the quasi-uniform state of the free layer magnetization is the lowest energy state in a significant range of fields near zero. The asymmetry of the total energy in $H_\mathrm{x}$ is due to a non-zero net magnetic moment of the SAF layer.

Micromagnetic simulations of the hysteresis loop at room temperature ($T \approx$ 300\,K) and magnetic field sweep rate employed in our experiment are prohibitively time consuming. However, given the importance of thermal fluctuations for a free layer nanomagnet of the small size and low magnetic anisotropy employed in our measurements, we can expect the free layer is in its lowest energy state for magnetic field values near zero. This assumption is supported by the absence of hysteresis in the experimentally measured $M_\mathrm{x}(H_\mathrm{x})$ curve shown in Fig.\,1b of the main text. Furthermore, the data in Supplementary Fig.\,1d show that the Boltzmann probability of the quasi-uniform state is much greater than that of the domain wall state in the middle of the hysteresis curve. For any two states, the Boltzmann ratio of probabilities of being in those states is: 
\begin{equation}
\frac{p_i}{p_j} = \exp{\left[-(E_i-E_j)/k_\mathrm{B} T\right]},
\end{equation}
where $p_i$ and $p_j$ are the probabilities of being in state $i$ and state $j$, $E_i$ and $E_j$ are the total micromagnetic energies of the states, $k_\mathrm{B}$ is the Boltzmann constant, and $T$ is the temperature. The energy difference between the domain wall state and the quasi-uniform state in the middle of the hysteresis loop is $1.19 \times 10^{-19}$\,J, with the quasi-uniform state being lower in energy as shown in Supplementary Fig.\,1d. Therefore, the Boltzmann probability of the domain wall state is $3.5 \times 10^{-13}$ of that of the quasi-uniform state at $T = 300$\,K.

\section{Gilbert damping}
The Gilbert damping of the free layer was estimated from the spectral linewidth of the quasi-uniform mode measured by ST-FMR technique at $H_\mathrm{x} = 0.06$\,kOe. Assuming uniaxial anisotropy, the Gilbert damping parameter is given by the ratio of half width at half maximum $\Delta f$ of the ST-FMR resonance curve $V_\mathrm{r}(f_\mathrm{d})$ to the quasi-uniform mode resonance frequency $f_\mathrm{SW}$ \cite{Gurevich1996}:  
\begin{equation}
\alpha = \frac{\Delta f}{f_\mathrm{SW}}.
\label{eq:gilbert}
\end{equation}
In the experiment (see Fig.\,1c of the main text), we find $\Delta f \approx$ 0.03\,GHz. By using this value of the linewidth and the measured resonance frequency $f_\mathrm{SW}=$ 0.91\,GHz, we obtain an estimate of the free layer Gilbert damping constant $\alpha \approx$ 0.033. 

\section{VCMA efficiency}
The observed linear shift of the quasi-uniform mode resonance frequency $f_\mathrm{SW}$ with applied direct voltage $V_\mathrm{dc}$ shown in Fig.\,1e of the main text arises exclusively from VCMA. The effective fields due to field-like and damping-like spin torque are perpendicular to the free layer magnetization for the perpendicular orientation of the free and the SAF magnetic moments employed in our experiment. Such perpendicular fields can only induce a quadratic shift of the quasi-uniform mode frequency. The frequency shift due to Ohmic heating is independent of the current polarity and thus is also quadratic in $V_\mathrm{dc}$ to leading order. Given the linear relation between the resonance frequency $f_\mathrm{SW}$ and the anisotropy field $H_\mathrm{u}$ for a uniaxial ferromagnet, the slope of the line in Fig.\,1e in the main text is  
\begin{equation}
\frac{\gamma}{2\pi} \frac{d H_\mathrm{u}}{d V_\mathrm{dc}},
\end{equation}
where the gyromagnetic ratio $\gamma$ is taken to be 176\,GHz/T. The data in Fig.\,1e in the main text gives VCMA efficiency $\frac{d H_\mathrm{u}}{d V_\mathrm{dc}}= 526$\,Oe/V, which is typical for this material system \cite{Zhu2012}.

\section{Theory of parametric resonance threshold}

For the theoretical description of parametric resonance of the MTJ free layer we employ a single-mode approximation. We expand the free layer magnetization into static and dynamic parts: $\mathbf{M}(\mathbf{r},t) = M_\mathrm{s}(\boldsymbol{\mu} + c(t)\mathbf{m}(\mathbf{r}) + c^*(t)\mathbf{m}^*(\mathbf{r}))$, where $\boldsymbol{\mu}$ is the unit vector in the direction of the static magnetization, $\mathbf{m}(\mathbf{r})$ is the coordinate-dependent vector structure of the spin wave mode, and $c$ is the dimensionless amplitude of this mode. Starting from the Landau-Lifshitz-Gilbert equation, the following nonlinear equation describing the dynamics can be derived \cite{Gurevich1996,Verba2014}:
\begin{equation}
\label{eq:th1}
\frac{\mathrm{d}c}{\mathrm{d}t} + i (\omega_\mathrm{SW} + \Psi |c|^2) c + \Gamma c = h V_{00} e^{i \omega_\mathrm{p} t} c^* + \eta(t),
\end{equation}
where $\omega_\mathrm{SW} = 2 \pi f_\mathrm{SW}$ is the spin wave mode angular frequency, $\Psi$ is the nonlinear frequency shift of the mode, $\Gamma$ is the damping rate of the mode, $h$ is the effective pumping field amplitude, $\omega_\mathrm{p}$ is the pumping frequency, $V_{00}$ is the efficiency of parametric interaction, and $\eta(t)$ describes thermal noise (see Ref.~\onlinecite{Verba2013} for details). In these notations, the parametric resonance threshold field is $h_\mathrm{th} = \Gamma / |V_{00}|$,  where $|V_{00}| = \frac{\gamma \mu_0}{2} \varepsilon$, $\gamma$ is the gyromagnetic ratio taken to be 176\,GHz/T, $\mu_0= 4 \pi \times 10^{-7}$\,T$\cdot$m/A is the permeability of vacuum, and the damping rate $\Gamma = 2 \pi \Delta f = 2 \pi \times 0.03$\,GHz \cite{Gurevich1996,Verba2014}. Averaged ellipticity of the spin wave mode $\varepsilon$ is given by \cite{Verba2014}:
\begin{equation}
\varepsilon = \left| \frac{\left<\mathbf{m}^*\cdot\mathbf{m}^*\right>_\mathbf{r}}{\left<\mathbf{m}^*\cdot(\boldsymbol{\mu}\times\mathbf{m})\right>_\mathbf{r}} \right|, 
\end{equation}
where $\left<...\right>_\mathbf{r}$ denotes a spatial average over the free layer volume. 

We calculate the quasi-uniform mode ellipticity from the micromagnetic mode profile. This calculation gives $\varepsilon = 0.26$, which results in $h_\mathrm{th}$ = 6.5\,kA/m = 82\,Oe. This gives the threshold voltage for excitation of parametric resonance $V_\mathrm{th} = h_\mathrm{th} \frac{d V_\mathrm{dc}}{d H_\mathrm{u}}$ = 0.156\,V. This micromagnetic value of $V_\mathrm{th}$ is significantly higher than that given by the macrospin approximation with $\varepsilon = \omega_\mathrm{M} |N_\mathrm{x} - N_\mathrm{y} |/ (2 \omega_\mathrm{SW})$, where $\omega_\mathrm{M} = \gamma \mu_0 M_\mathrm{s}$ (with $M_\mathrm{s}~=~950$\,kA/m), $N_\mathrm{x} = 0.014$ and $N_\mathrm{y} = 0.040$ are components of the free layer demagnetization tensor \cite{Beleggia2005}, and $\omega_\mathrm{SW} = 2 \pi \times 0.91$\,GHz is the spin wave mode frequency. The higher value of ellipticity ($\varepsilon = 0.478$) in the macrospin approximation leads to a lower parametric threshold: $h_\mathrm{th}$ = 45\,Oe and $V_\mathrm{th}$ = 0.085\,V. As expected, the experimentally measured value of the threshold voltage $V_\mathrm{th}$ = 0.136\,V is much higher than that given by the macrospin approximation but it is similar to that appropriate for micromagnetic profile of the quasi-uniform mode. The 15\% discrepancy between the measured and the theoretically predicted threshold could arise from deviation of the free layer shape from the ideal elliptical shape assumed in the simulations and from over-estimation of the damping parameter of the free layer. 

\section{Evaluation of the parametric resonance threshold from experiment}

To determine the threshold voltage for parametric excitation $V_\mathrm{th}$ from the experimental data in Fig.\,3 of the main text, we fit these data to theoretical expressions of the oscillation power as a function of the drive amplitude $V_\mathrm{ac}$. These expressions are derived below for two limits: $V_\mathrm{ac}\ll V_\mathrm{th}$ and $V_\mathrm{ac}\gg V_\mathrm{th}$.  

\subsection{Below the threshold}

Well below the threshold ($V_\mathrm{ac}\ll V_\mathrm{th}$), the nonlinear frequency shift in Eq.\,\eqref{eq:th1} can be neglected and the parametric resonance frequency $f_\mathrm{pr}$ is equal to twice the spin wave mode frequency ($f_\mathrm{pr}=2f_\mathrm{SW}$). In this limit, the reduced integrated power $p$ of the spin wave mode is given by the expression below when the free layer is driven exactly at the parametric resonance frequency $f_\mathrm{pr}$: 
\begin{equation}
\label{eq:intp_low}
p = \left<|c|^2\right> = \frac{C_1}{\Gamma - |h V_{00}|} + \frac{C_1}{\Gamma + |h V_{00}|} = \frac{D}{(V_\mathrm{th}-V_\mathrm{ac})} + \frac{D}{ (V_\mathrm{th}+V_\mathrm{ac})}, 
\end{equation}
where $\left<...\right>$ denotes a thermal average. In deriving this expression, we assumed white thermal noise: $\left<\eta(t)\eta(\tau)\right> = C_2 \delta(t - \tau)$ and $\left<\eta(t) \eta^*(\tau)\right> = C_1 \delta(t - \tau)$, where $C_1$, $C_2$ and $D=C_1 V_\mathrm{th}/\Gamma$ are constants. Here we also employed the linear relation between the effective (VCMA) pumping field amplitude $h$ and the microwave voltage amplitude $V_\mathrm{ac}$, which is evident from Fig.\,1e of the main text.

By expanding the noise term into Fourier series, we obtain the following expression for reduced power spectral density $p(f)$ of the spin wave mode oscillations:
\begin{eqnarray}
p(f)=\left<|c(f)|^2\right> = \frac{C_2}{(\Gamma - |h V_{00}|)^2 + (2\pi(f - f_\mathrm{SW}))^2} + \frac{C_2}{(\Gamma + |h V_{00}|)^2 + (2\pi(f - f_\mathrm{SW}))^2} \\ 
=\frac{A}{(V_\mathrm{th}-V_\mathrm{ac})^2 + (2\pi(f - f_\mathrm{SW}) V_\mathrm{th}/\Gamma)^2} + \frac{A}{(V_\mathrm{th}+V_\mathrm{ac})^2 + (2\pi(f - f_\mathrm{SW}) V_\mathrm{th}/\Gamma)^2}, 
\label{eq:peakp_low}
\end{eqnarray}
where $A=C_2 V_\mathrm{th}^2/\Gamma^2$ is a constant. Setting $f=f_\mathrm{SW}$ in Eq.\,\eqref{eq:peakp_low}, we obtain an expression for the peak value of the reduced PSD that is observed at $f= f_\mathrm{SW} = f_\mathrm{pr}/2$:
\begin{equation}
p(f_\mathrm{pr}/2)=\left<|c(f_\mathrm{pr}/2)|^2\right>  = \frac{A}{(V_\mathrm{th}-V_\mathrm{ac})^2} + \frac{A}{(V_\mathrm{th}+V_\mathrm{ac})^2}. 
\label{eq:psd}
\end{equation}

The second term in Eq.\,\eqref{eq:psd} is much smaller than the first one for $V_\mathrm{ac}$ approaching $V_\mathrm{th}$ and it can be neglected in fitting the experimental data of Fig.\,3 of the main text:

\begin{equation}
p(f_\mathrm{pr}/2)=\left<|c(f_\mathrm{pr}/2)|^2\right>  = \frac{A}{(V_\mathrm{th}-V_\mathrm{ac})^2}. 
\label{eq:psd_simple}
\end{equation}

\subsection{Above the threshold}
Well above the threshold ($V_\mathrm{ac}\gg V_\mathrm{th}$), the integrated power of the parametrically excited quasi-uniform mode is nearly temperature independent and can be approximated by its zero-temperature value \cite{Slavin2009}.  Neglecting the thermal noise term in Eq.\,\eqref{eq:th1}, we derive:
\begin{equation}
p=|c|^2 = \frac{1}{|\Psi|}\left(\sqrt{(h V_{00})^2 - (h_\mathrm{th} V_{00})^2} + 2\pi(f_\mathrm{pr}/2-f_\mathrm{SW}) \textrm{sign}(\Psi)\right).
\label{eq:above_th}
\end{equation}

We assume that the deviation of $f_\mathrm{pr}$ from $2f_\mathrm{SW}$ is small, so that the second term in the parentheses can be neglected compared to the first term. In this case, Eq.\,\eqref{eq:above_th} takes a simple form:
\begin{equation}
p=B\sqrt{V_\mathrm{ac}^2-V_\mathrm{th}^2},
\label{eq:above_th_simple}
\end{equation}
where $B= \Gamma /(|\Psi| V_\mathrm{th})$ is a constant. 

\subsection{Details of the fitting procedure}

The fitting of the normalized peak power data shown in Fig.\,3 of the main text to Eq.\,\eqref{eq:psd_simple} and Eq.\,\eqref{eq:above_th_simple} was performed by the least squares method with $A$, $B$ and $V_\mathrm{th}$ as fitting parameters. A range of data near the threshold voltage must be excluded in the fitting procedure because neither Eq.\,\eqref{eq:psd_simple} nor Eq.\,\eqref{eq:above_th_simple} is valid at the threshold voltage. We chose the data range where $V_\mathrm{ac} < 0.1$\,V for the low-power fit and the data range where $V_\mathrm{ac} > 0.16$\,V for the high-power fit (the excluded data range is $0.1$\,V -- $0.16$\,V) because the best fit parameters do not change significantly upon further extension of the excluded data range. The threshold voltage given by this fitting procedure is $V_\mathrm{th} = 0.136$\,V.

\end{document}